\documentclass[%
 aip,amsmath,amssymb,
 reprint,%
]{revtex4-1}
\usepackage{amsmath}

\usepackage{graphicx}% Include figure files
\usepackage{dcolumn}% Align table columns on decimal point
\usepackage{bm}% bold math

\newcommand{\kT}{k_{\rm B}T}
\newcommand{\md}{\mathrm{d}}
\usepackage{xcolor} %color stuff
\usepackage[normalem]{ulem} %strike through text
\newcommand{\ML}{}
\newcommand{\stkout}[1]{}

\draft 

\begin{document}

\title{Multidimensional minimum-work control of a 2D Ising model} 

\author{Miranda D.\ Louwerse}
\email{mdlouwer@sfu.ca}
\affiliation{%
 Department of Chemistry, Simon Fraser University, Burnaby, British Columbia, Canada V5A1S6
}%
\author{David A.\ Sivak}%
 \email{dsivak@sfu.ca}
\affiliation{%
 Department of Physics, Simon Fraser University, Burnaby, British Columbia, Canada V5A1S6
}%

\date{\today}

%Please provide a brief explanation of the significance and originality of your contribution as well as how it advances the field:
%We undertook a detailed investigation of multidimensional protocols designed to minimize work for a simple model system, revealing potential design principles of efficient control in multiple dimensions that qualitatively differ from principles of efficient one-dimensional control that have been deduced in previous works. Therefore our results are a step towards a more complete understanding of energetically efficient driving, which in turns provides insight into the efficient operation of biomolecular machines and has implications for free-energy estimation. Additionally, we demonstrate several advantages and interesting features that arise from efficient multidimensional driving that could generalize to a wide variety of systems. 

\begin{abstract}
A system's configurational state can be manipulated using dynamic variation of control parameters, such as temperature, pressure, or magnetic field; for finite-duration driving, excess work is required above the equilibrium free-energy change. Minimum-work protocols in multidimensional control-parameter space have the potential to significantly reduce work relative to one-dimensional control. By numerically minimizing a linear-response approximation to the excess work, we design protocols in control-parameter spaces of a 2D Ising model that efficiently drive the system from the all-down to all-up configuration. We find that such designed multidimensional protocols take advantage of more flexible control to avoid control-parameter regions of high system resistance, heterogeneously input and extract work to make use of system relaxation, and flatten the energy landscape, making accessible many configurations that would otherwise have prohibitively high energy and thus decreasing spin correlations. Relative to one-dimensional protocols, this speeds up the rate-limiting spin-inversion reaction, thereby keeping the system significantly closer to equilibrium for a wide range of protocol durations, and significantly reducing resistance and hence work. 
\end{abstract}


\maketitle 

\section{\label{sec:Introduction} Introduction}

Driving a stochastic system between configurations using time-dependent external control parameters (e.g., pressure, temperature, or magnetic field) is a central tool in single-molecule experiment and simulation. While quasistatic (infinitely slow) protocols require the same work (equal to the equilibrium free-energy difference between control-parameter endpoints) independent of the protocol choice, finite-duration protocols drive the system out of equilibrium and accrue varying amounts of excess work above the equilibrium free-energy change. Since practical applications require finite-duration driving, there is significant motivation to improve our understanding of protocols that minimize work~\cite{Crooks2007,Sivak2012,Schmiedl2007,Large2019}, which can in turn provide insight into energetically efficient operation of biological molecular machines~\cite{Brown2020} and allow more efficient estimation of equilibrium system properties~\cite{Shenfeld2009,Kim2012,Blaber2020}. For example, the free-energy difference between equilibrium ensembles or the free-energy profile along the driving coordinate can be estimated from system response to control protocols using nonequilibrium work relations~\cite{Jarzynski1997,Crooks2000, Hummer2001}, and the precision of free-energy estimates depends on the protocol excess work~\cite{Kim2012,Blaber2020}. Control protocols can be performed in simulation~\cite{Dellago2014,Morfill2008} or experiment~\cite{Liphardt2002,Tafoya2019,Gupta2011,Engel2014}, providing access to molecular-level information about a wide variety of systems.

When the protocol is sufficiently slow, excess work can be approximated via linear-response theory~\cite{Sivak2012}, yielding a Riemannian metric in control-parameter space measuring the excess power during the protocol. The metric is interpreted as a generalized friction tensor, quantifying system resistance (due to equilibrium fluctuations) to changes in the control parameter. The friction encodes information about changes in the system’s equilibrium free-energy during the protocol~\cite{Crooks2007} (its thermodynamics) and equilibrium relaxation modes~\cite{Bonanca2014,Mandal2016} (its kinetics). A general design principle resulting from this theory is an inverse relationship between control-parameter velocity and friction: minimum-work protocols reduce driving speed where the friction is large, allowing time for the system to relax due to stochastic fluctuations that do not require work~\cite{Sivak2016}.

Many systems of interest have slow relaxations between metastable mesostates, and finding a good control parameter to drive the system can be non-trivial~\cite{Moradi2014}. For finite-duration protocols where the system has insufficient time to fully relax, the system can get stuck in metastable mesostates, causing the nonequilibrium system distribution to lag the equilibrium distribution, resulting in excess work. A longer-duration protocol would allow further time for the system to relax from these metastable mesostates toward the equilibrium distribution and reduce excess work, but it may not be feasible or desirable to increase duration.

A similar problem arises in enhanced-sampling contexts, where external control parameters are used to bias system degrees of freedom that are relevant to characterizing metastable mesostates, reducing the time required to sample the system space and compute free energies~\cite{Tiwary2016,Yang2019}. One seeks a collective variable that captures the slowest relaxation mode, otherwise the system can get stuck in a metastable region, resulting in slow convergence of relevant statistics (e.g., free-energy profiles)~\cite{Moradi2014,Morfill2008}. Using multiple biasing coordinates that couple to relevant relaxation modes can significantly improve the sampling speed by allowing the system to more easily circumvent dynamical barriers~\cite{Maragliano2006a,Pfaendtner2015,Jiang2012,Zhao2017,Chipot2011}.

Thus, there is interest in understanding how increasing the number of control parameters influences the design of minimum-work protocols~\cite{Dellago2014}, both for the thermodynamic benefit of reducing work and the dynamic benefit of speeding system relaxation (allowing shorter-duration protocols). Here, we design protocols to minimize work in a multidimensional control-parameter space of a 3$\times$3 Ising model (Fig.~\ref{fig:Ising_Model}), driving the system to invert its magnetization from all-down to all-up using an external magnetic field. We contrast a single magnetic field applied to all spins (driving the total system magnetization) with a set of fields each applied independently to a subset of spins. These designed multidimensional protocols avoid high friction and decorrelate spins by flattening the energy landscape, allowing the system to access configurations with high internal energy and significantly speeding system relaxation. This keeps the system closer to equilibrium and thereby reduces work relative to one-dimensional protocols. Overall, using multiple control parameters to drive the system offers greater flexibility in speeding up slow dynamical modes that prevent system relaxation to equilibrium; work can be done individually on each controlled degree of freedom and timed to take advantage of expected system relaxation to reduce resistance to driving.

\section{\label{sec:ModelSystem} Ising System}

\begin{figure}[t]
    \centering
    \includegraphics[width=\linewidth]{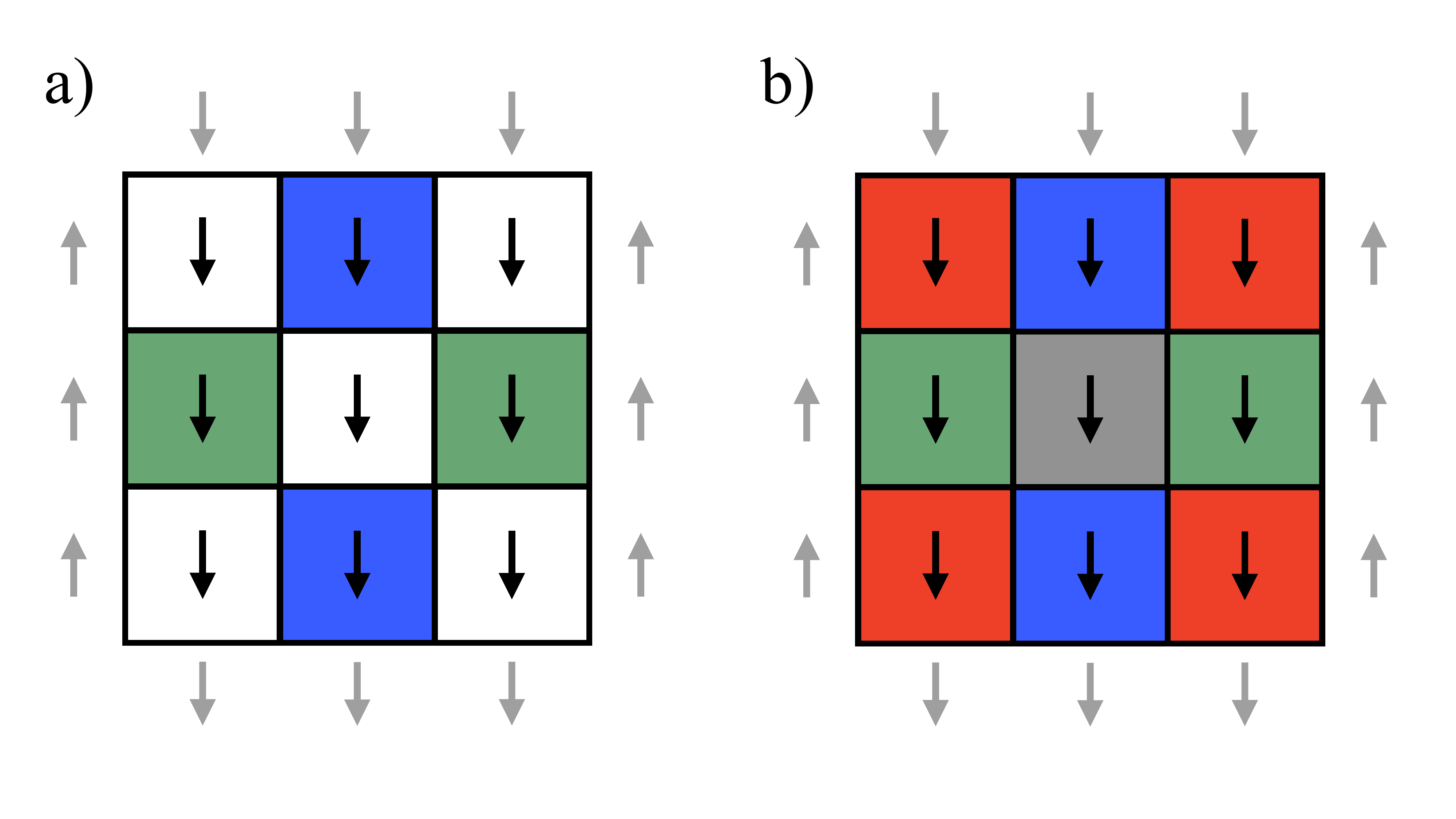}
    \caption{Schematic of the 3$\times$3 Ising model with nine fluctuating spins (black, shown in initial all-down configuration) and 12 fixed boundary spins (gray). (a) 2D control parameter given by two external fields on spins colored blue and green, respectively. (b) 4D control parameter given by four external fields on spins colored according to their symmetry type. Spins with the same color are influenced by a common magnetic field.}
    \label{fig:Ising_Model}
\end{figure}

Our system is a two-dimensional 3$\times$3 Ising model with fixed anti-symmetric boundary conditions~\cite{Rotskoff2017}, as illustrated in Fig.~\ref{fig:Ising_Model}. The spins are ferromagnetically coupled, with spin configuration $\{ \bm{\sigma} \}$ having internal energy
\begin{equation} \label{eq:system_energy}
    E_{\rm int}(\{ \bm{\sigma} \}) \equiv -J \sum_{\{i,j\}} \sigma_i \sigma_j \>,
\end{equation}
where $J=1\,\kT$ is the coupling coefficient for Boltzmann constant $k_{\rm B}$ and temperature $T$, $\sigma_i \in \{ -1, 1 \}$ is the orientation of spin $i$, and $\sum_{\{i,j\}}$ denotes a sum over nearest-neighbor spin pairs. The system evolves under single-spin-flip Glauber dynamics~\cite{Glauber1963} and has two energetically metastable configurations, with spins all down or all up.
{\ML The dynamical barrier separating the metastable configurations has an activation energy $\Delta E^{\ddagger} = 8 \kT$; the spin-inversion transition from all down to all up is a relatively rare event, with the mean first-passage time of $\sim2 \times 10^3$ attempted spin flips~\cite{Vanden-Eijnden2014}.}

We drive the spin-inversion reaction to flip the system configuration from all-down to all-up using a vector of external magnetic fields $\boldsymbol{h}$ applied to sets of spins (background colors in Fig.~\ref{fig:Ising_Model}), chosen to reflect the symmetry of the boundary conditions. The applied magnetic fields $\bm{h}$ bias the total energy as
\begin{equation} \label{eq:control_energy}
    E_{\rm tot}(\{ \bm{\sigma} \},\bm{h}) \equiv E_{\rm int}(\{ \bm{\sigma} \})-\bm{h}^{\rm T} \cdot \bm{X}(\{ \bm{\sigma} \}) \>,
\end{equation}
where $\bm{X}(\{ \bm{\sigma} \})$ is a vector of conjugate forces for configuration $\{ \bm{\sigma} \}$, with force $X_i \equiv -\partial E_{\rm tot} / \partial h_i$ conjugate to control parameter (external field) $h_i$, equaling the total magnetization of the spin set controlled by $h_i$. We invert the system magnetization by changing the magnetic field from $h_i(t=0)=-0.5\,\kT$ for all $i$ (favoring the all-down configuration) to $h_i(t=\Delta t) = 0.5\,\kT$ (favoring the all-up configuration).

We explore two multidimensional control-parameter spaces associated with a 2D magnetic field applied to the blue and green spins (Fig.~\ref{fig:Ising_Model}(a), the spins that are most biased by the boundary conditions) or a 4D magnetic field applied to all spins [Fig.~\ref{fig:Ising_Model}(b), the highest dimensionality that does not break the symmetry imposed by the boundary conditions]. The 2D control-parameter space is a two-dimensional manifold of the 4D control-parameter space with $h_{\rm red} = h_{\rm black} = 0 \kT$.

\section{\label{sec:MWCP_Theory} Linear-response approximation of excess work}

With the system initially at equilibrium at control parameter $\bm{h}(t=0)$, we drive the system in finite duration $\Delta t$ between control-parameter endpoints $\bm{h}(t=0)$ and $\bm{h}(t=\Delta t)$ following a prescribed protocol of changes to the magnetic fields. We decompose the mean work performed on the system during the protocol into quasistatic and excess work,
\begin{equation} \label{eq:protocol_work_qs_ex}
    \mathcal{W} = \mathcal{W}^{\rm qs} + \mathcal{W}^{\rm ex} \>.
\end{equation}
The quasistatic work is the work that would be performed during an infinitely long protocol where the system constantly relaxes to remain in equilibrium with the current control parameters, and it equals the difference in free energy $F(\bm{h})=-\kT \ln \sum_{\{ \bm{\sigma} \}} e^{-\beta E_{\rm tot}(\{ \bm{\sigma} \},\bm{h})}$ between equilibrium distributions corresponding to the control-protocol endpoints, $\mathcal{W}^{\rm qs} = \Delta F = F(\bm{h}(t=\Delta t))-F(\bm{h}(t=0))$. $\beta=1/\kT$ is the inverse temperature. When the protocol is performed in finite duration, the system is driven out of equilibrium, and positive excess work $\mathcal{W}^{\rm ex}$ is done which is dissipated as heat as the system relaxes to equilibrium during and after completion of the protocol.

For sufficiently long (yet finite-duration) protocols, the mean excess work over a protocol can be approximated via linear-response theory as~\cite{Sivak2012}
\begin{equation} \label{eq:excess_work}
    \mathcal{W}^{\rm ex} \approx \int_0^{\Delta t} \md t \ \sum_{i,j} \dot{h}_i(t) \zeta_{ij}(\boldsymbol{h}(t)) \dot{h}_j(t) \ ,
\end{equation}
for control-parameter velocities $\dot{h}_i(t)$ and friction tensor
\begin{equation} \label{eq:friction_tensor}
    \zeta_{ij}(\bm{h}(t)) \equiv \beta \int_0^{\infty} \md t' \ \langle \delta X_j(0) \delta X_i(t') \rangle_{\bm{h}(t)}^{\rm eq} \ ,
\end{equation}
the time integral of the temporal correlation function $\langle \delta X_j(0) \delta X_i(t') \rangle_{\bm{h}(t)}^{\rm eq}$ of forces $X_i$ and $X_j$ respectively conjugate to control parameters $h_i$ and $h_j$. The correlation function is for the system at equilibrium at control parameter $\bm{h}(t)$, and $\delta X_i(t') \equiv X_i(t')-\langle X_i \rangle_{\bm{h}(t)}^{\rm eq}$ is the difference of the instantaneous conjugate force from its equilibrium mean. The friction can be factored into two components,
\begin{equation} \label{eq:friction_decomposition}
    \zeta_{ij}(\bm{h}(t)) = \beta \langle \delta X_j \delta X_i \rangle_{\bm{h}(t)}^{\rm eq} \tau_{ij}(\bm{h}(t))  \ ,
\end{equation}
the equilibrium covariance $\langle \delta X_j \delta X_i \rangle_{\bm{h}(t)}^{\rm eq}$ and the integral relaxation time $\tau_{ij}(\bm{h}(t))=\int_0^{\infty} \md t' \ \frac{\langle \delta X_j(0) \delta X_i(t') \rangle_{\bm{h}(t)}^{\rm eq}}{\langle \delta X_j \delta X_i \rangle_{\bm{h}(t)}^{\rm eq}}$ between conjugate forces $X_i$ and $X_j$.

The friction tensor defines a Riemannian metric in control-parameter space, where minimum-work protocols are geodesics (shortest paths) between chosen endpoints~\cite{Crooks2007}. This also implies that during the minimum-work protocol the excess power (the integrand of Eq.~\eqref{eq:excess_work}) is constant.

\section{\label{sec:Methods} Methods}
In both 2D and 4D control-parameter spaces we consider three types of protocols between the same endpoints. A naive protocol changes all magnetic fields with constant velocity. The time-optimized protocol changes all magnetic fields together (i.e., $h_i(t)=h_j(t)$ for all $i,j,t$), with velocity optimized to minimize (within the linear-response approximation) the work. 
Both these protocol types are equivalent to one-dimensional control using a single magnetic field to drive the system, with conjugate force the total magnetization of controlled spins (only blue and green spins' magnetization for 2D, and total magnetization for 4D). This yields a one-dimensional friction coefficient, the sum of all elements of the 2D or 4D friction matrices $\zeta^{\rm 1D}(\bm{h}) = \sum_{ij} \zeta_{ij}(\bm{h})$. During the time-optimized protocols, all fields are changed with velocity proportional to the inverse square-root of the one-dimensional friction coefficient~\cite{Sivak2012}, $\dot{h}^{\rm 1D} \propto \zeta^{\rm 1D}(\bm{h})^{-1/2}$. The fully optimized protocols solve the Euler-Lagrange equation~\cite{Morin2008} for Eq.~\eqref{eq:excess_work} in all control-parameter dimensions, which is done numerically using the string method~\cite{Rotskoff2017}. 

We estimated the 4$\times$4 friction tensor~\eqref{eq:friction_tensor} on a discrete grid in 4D control-parameter space and used this friction landscape to design fully optimized 2D and 4D protocols. Sampling the friction in a multidimensional space has a significant computational cost since the number of grid points scales exponentially with the number of control parameters; in contrast, the time-optimized protocol can be calculated by estimating the friction tensor along the one-dimensional protocol path, which has a significantly smaller computational cost. (The naive protocol requires no prior sampling.) We simulate each protocol type for a range of protocol durations and collect an ensemble of system responses. Appendix~\ref{computational_details} provides details on numerical methods. 

Throughout, we report work and power divided by the number $N_i$ of spins controlled by field $i$ ($N_{\rm red}=4$, $N_{\rm blue}=N_{\rm green}=2$, and $N_{\rm black}=1$). We also report the protocol duration scaled by a reference relaxation time $\tau_{\rm rel}$, the estimated protocol duration required for on average one all-down to all-up transition to occur (see Appendix~\ref{reference_relaxation_time} for details).

\section{\label{sec:Results} Results}

\subsection{Avoiding high friction}

\begin{figure}[ht]
    \centering 
    \includegraphics[width=\linewidth]{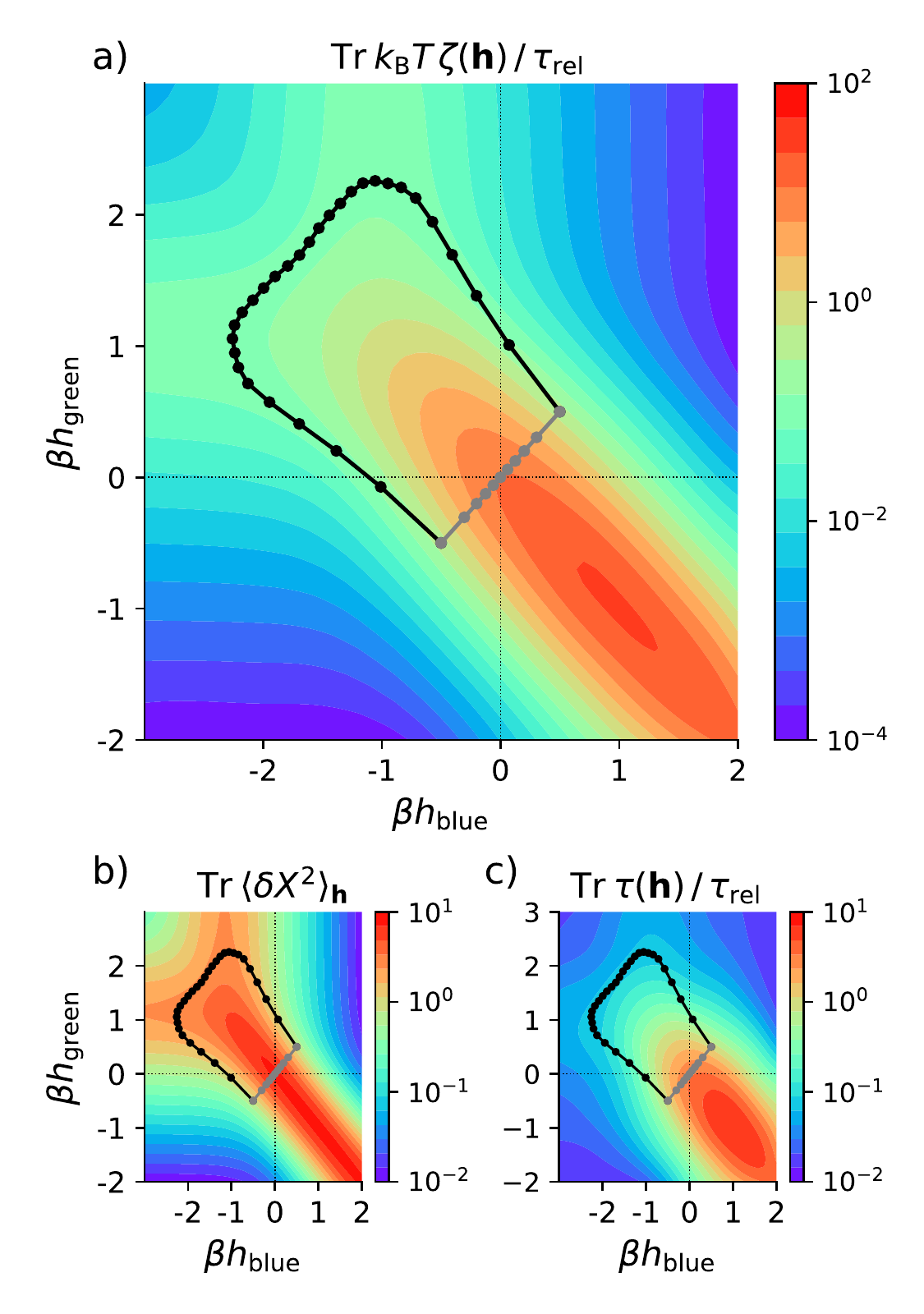} 
    \caption{2D fully optimized (black) and time-optimized (gray) protocols overlaid on a heatmap showing the trace of the (a) friction~\eqref{eq:friction_tensor}, (b) force covariance, or (c) integral relaxation time~\eqref{eq:friction_decomposition} matrix. Points denote control-parameter values evenly spaced in time on the given protocol.}
    \label{fig:2D_MWCP_friction_surface}
\end{figure} 

To illustrate general features of the multidimensional friction tensor for this system, Fig.~\ref{fig:2D_MWCP_friction_surface}(a) shows the trace (a scalar capturing the essential features) of the 2$\times$2 friction tensor in the control-parameter space defined by axes $(h_{\rm blue}, h_{\rm green})$. The friction is maximized at $(1,-1)$, where the fields cancel the fixed boundary conditions (Fig.~\ref{fig:Ising_Model}), stabilizing the all-down and all-up configurations relative to the unperturbed system ($h_i = 0$ for all $i$). Both components of the friction~\eqref{eq:friction_decomposition}, the force covariance [Fig.~\ref{fig:2D_MWCP_friction_surface}(b)] and the integral relaxation time [Fig.~\ref{fig:2D_MWCP_friction_surface}(c)], are similarly peaked.  

Figure~\ref{fig:2D_MWCP_friction_surface} also shows the 2D time-optimized and fully optimized protocols. (The naive protocol follows the same path as the time-optimized protocol but with constant protocol velocity.) The naive and time-optimized protocols follow a straight line between the fixed endpoints and drive the system through a high-friction region, with the time-optimized protocol slowing down where the friction along the path is large, near control parameters corresponding to the unperturbed system. In contrast, the fully optimized protocol avoids this by making an excursion through low-friction regions of control-parameter space. 

\begin{figure}[ht]
    \centering
    \includegraphics[width=\linewidth]{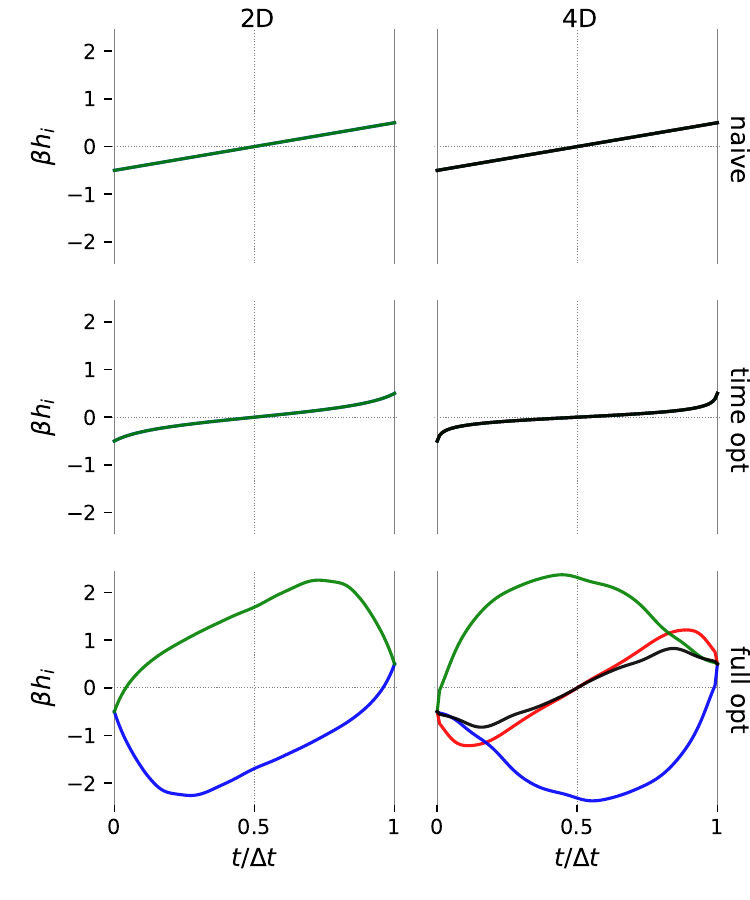} 
    \caption{Designed protocols between fixed endpoints. Magnetic fields $h_i$ as a function of scaled protocol time $t/\Delta t$ during naive (top row), time-optimized (middle), and fully optimized (bottom) protocols in 2D (left column) and 4D (right) control-parameter spaces. Colors correspond to spin sets in Fig.~\ref{fig:Ising_Model}.}
    \label{fig:protocol_paths}
\end{figure} 

Figure~\ref{fig:protocol_paths} shows the three protocol types in 2D and 4D control-parameter spaces. The time-optimized protocols have high initial and final velocity, and low velocity in the middle of the protocol, where fields are close to zero and friction is large. This is consistent with the minimum-work protocol for one-dimensional barrier crossing~\cite{Sivak2016}, where the protocol slows to allow time for stochastic fluctuations to kick the system over an energy barrier, thereby remaining closer to equilibrium relative to the naive protocol.

The fully (both temporally and spatially) optimized protocols depart dramatically from the naive and time-optimized protocols. Consistent with the results in a larger Ising model~\cite{Rotskoff2017}, the green field (whose corresponding spins are initially anti-aligned with their boundary conditions and therefore energetically frustrated) is increased rapidly in the early stages of the protocol, while the blue field (whose corresponding spins are initially aligned with their boundary conditions) is decreased initially and then increased later. 2D and 4D protocols show the same trend, but the addition of time-varying red and black fields in 4D protocols shifts the relative timing of changes in green and blue fields.

A striking feature of both the 2D and 4D fully optimized protocols is the non-monotonic magnetic fields at early and late stages. In the linear-response regime, a 1D minimum-work protocol must be monotonic since a non-monotonic protocol drives through the same control-parameter value more than once, producing unnecessary dissipation; in the metric language of the linear-response approximation, the shortest curve between two points cannot cross itself. However, individual fields for minimum-work protocols can be non-monotonic as long as the protocol path in multidimensional space does not loop back on itself.

\subsection{Reducing excess work and keeping close to equilibrium}

\begin{figure}[t] 
    \centering
    \includegraphics[width=\linewidth]{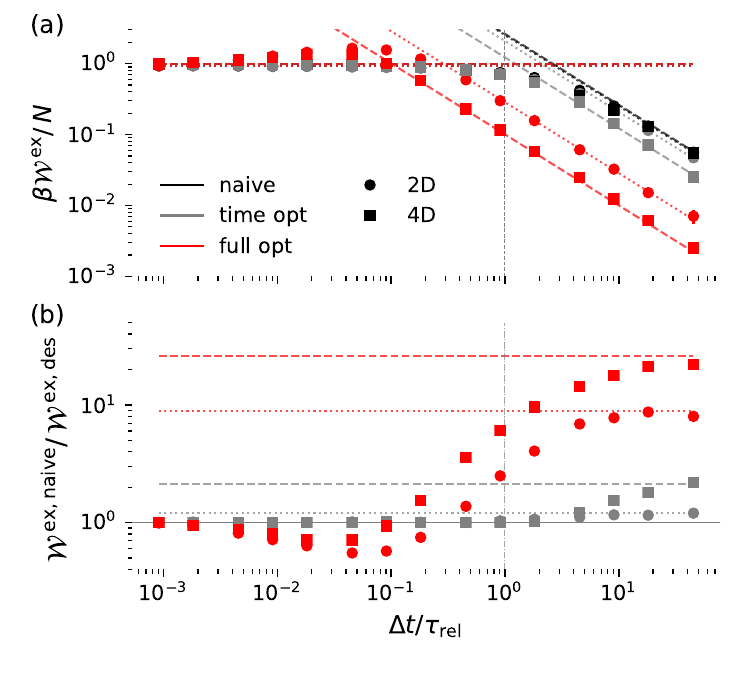}
    \caption{(a) Excess work~\eqref{eq:excess_work} scaled by the number $N$ of controlled spins (i.e., $N=4$ controlled spins for 2D protocols and $N=9$ for 4D) as a function of protocol duration $\Delta t$, for each of the six protocol types (Fig.~\ref{fig:protocol_paths}): 2D (circles) and 4D (squares) control-parameter spaces; and naive (black), time-optimized (gray), and fully optimized (red). (b) Ratio of excess work during the time-optimized (gray) and fully optimized (red) protocols to excess work in naive protocols for 2D (circles) and 4D (squares) control. Horizontal dotted lines: excess work for an instantaneous protocol. Dashed lines: linear-response approximations, accurate at long duration. Protocol durations are scaled by reference relaxation time $\tau_{\rm rel}$ (Appendix~\ref{reference_relaxation_time}).}
    \label{fig:workvsduration}
\end{figure}

We next compare the excess work for all protocols to assess their relative performance. Figure~\ref{fig:workvsduration}(a) shows the excess work as a function of protocol duration. In the limit of long duration, the excess works approach their respective linear-response approximations and scale as $\Delta t^{-1}$. For short protocols, the excess works approach that of an instantaneous protocol, given by the relative entropy of the initial and final equilibrium distributions~\cite{Large2019}. Figure~\ref{fig:workvsduration}(b) shows the ratio of excess work for naive relative to time-optimized and fully optimized protocols. While the 2D and 4D time-optimized protocols respectively do $\sim$1.2$\times$ and $\sim$2$\times$ less excess work than the corresponding naive protocols, the fully optimized protocols respectively do $\sim$9$\times$ and $\sim$26$\times$ less excess work than the naive protocols. [Fully optimized control protocols do more work than all other protocols (including an instantaneous protocol) for short durations ($\Delta t/\tau_{\rm rel} \lesssim 0.1$).]

\begin{figure}[ht]
    \centering
    \includegraphics[width=\linewidth]{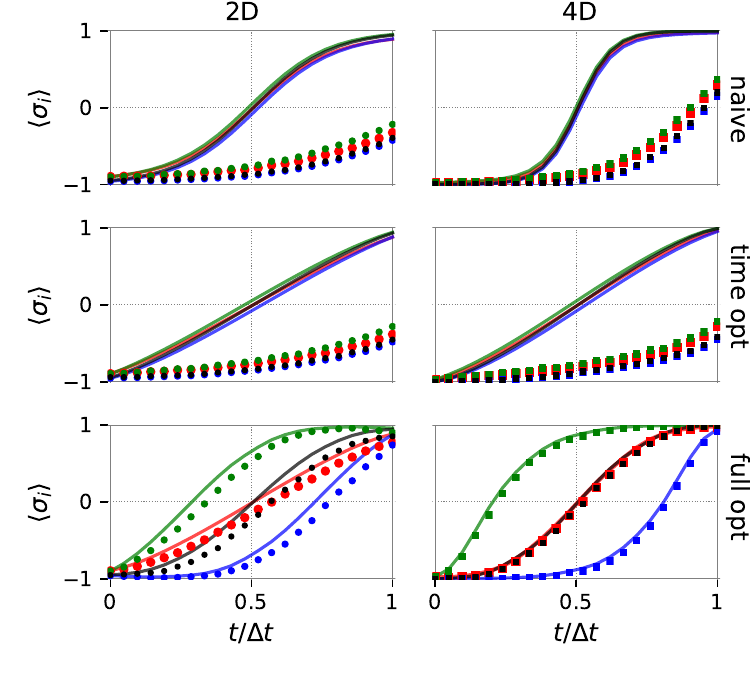}
    \caption{Quasistatic (curves) and nonequilibrium (points) mean magnetization of each spin type (different colors, see Fig.~\ref{fig:Ising_Model}) as a function of elapsed time $t$ along a moderate-duration ($\Delta t \approx \tau_{\mathrm{ref}}$) protocol of each type (see Fig.~\ref{fig:protocol_paths}).} 
    \label{fig:protocol_mean_trajectories}
\end{figure}

\begin{figure*}
	\centering
	\includegraphics[width=\linewidth]{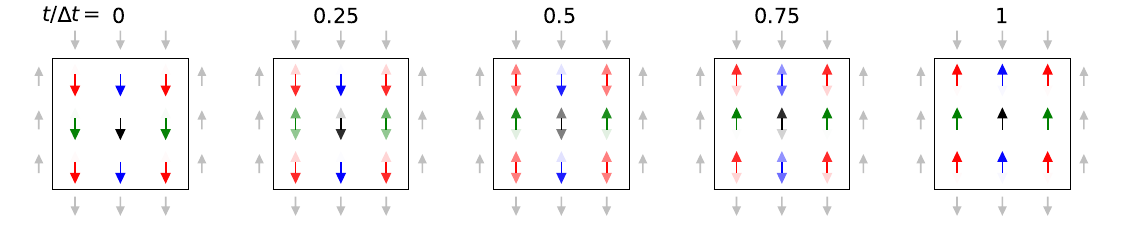}
	\caption{{\ML Spin-inversion mechanism at five evenly spaced times during the 4D fully optimized protocol. The transparency of up and down arrows indicate the equilibrium probability of each orientation.}}
	\label{fig:spin_inversion_mechanism}
\end{figure*}

We now show that fully optimized protocols keep the system much closer to equilibrium, resulting in significant work reduction. Figure~\ref{fig:protocol_mean_trajectories} shows for moderate protocol duration ($\Delta t \approx \tau_{\rm rel}$) the quasistatic and nonequilibrium mean magnetizations $\langle \sigma_i \rangle$ of each spin type. In fully optimized protocols, there is a rough ordering of spin flips matching the corresponding field increases (Fig.~\ref{fig:protocol_paths}): green spins flip first, red and black spins flip next, and blue spins flip last{\ML (also depicted schematically in 
Fig.~\ref{fig:spin_inversion_mechanism})}. This temporal separation of spin flips contrasts with naive and time-optimized protocols, where the different spin types have nearly equal mean magnetization throughout the protocol. 

Throughout the naive and time-optimized protocols, the system remains bistable, with the external fields primarily biasing the relative energies of the all-down and all-up configurations. For the moderate duration shown, the system gets stuck in the initial metastable basin so nonequilibrium mean magnetizations significantly lag the quasistatic limit. In contrast, the fully optimized protocols keep the system much closer to equilibrium, with modest lag during the 2D protocol and nearly no lag during the 4D protocol.

\subsection{Reducing spin covariance and flattening the energy landscape} 

We now turn to understanding how fully optimized protocols drive the system differently than naive protocols, extracting general features that are associated with reduced protocol work. 

\begin{figure}[t]
    \centering
    \includegraphics[width=\linewidth]{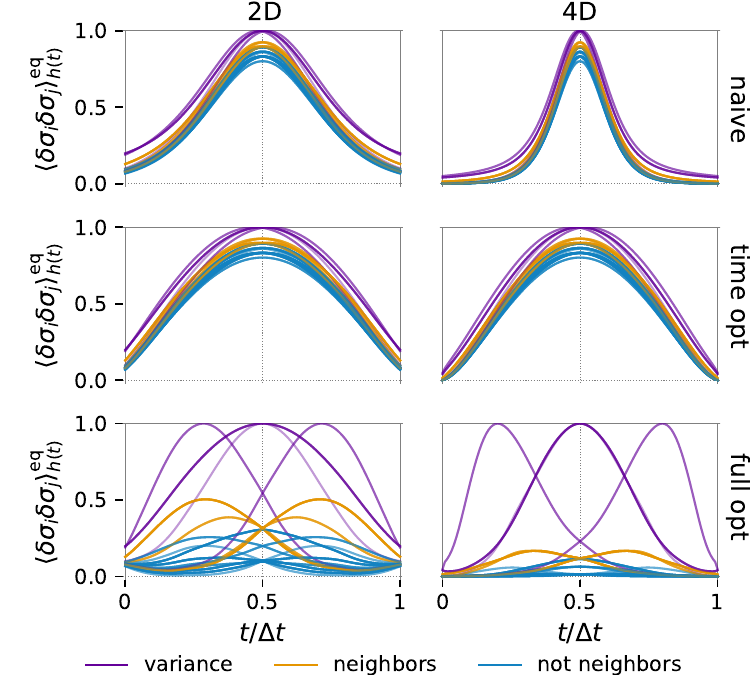}
    \caption{Equilibrium spin covariance as a function of elapsed time $t$ along a protocol of each type (Fig.~\ref{fig:protocol_paths}). Purple: spin variance. Yellow: covariance between neighboring spins. Blue: covariance between non-neighboring spins.}
    \label{fig:spin_covariance}
\end{figure}

Figure~\ref{fig:spin_covariance} shows the equilibrium variance of each spin and covariance between neighboring and non-neighboring spins. The 9$\times$9 spin-covariance matrix is ``coarse-grained'' (by summing spin-covariance elements for the same spin types into a single entry) to form the 4$\times$4 force-covariance matrix, a factor in the friction~\eqref{eq:friction_decomposition} that contributes to excess power for 4D protocols. Since each spin magnetization is a Bernoulli random variable, its is a quadratic function of the mean, $\langle \delta \sigma_i^2 \rangle=1-\langle \sigma_i \rangle^2$, where each variance peaks when its mean changes sign (Fig.~\ref{fig:protocol_mean_trajectories}). At some point during a long-duration protocol, the mean magnetization of each spin must change sign and hence the corresponding diagonal element of the force-covariance matrix must reach a fixed maximal value; however, the contribution from off-diagonal elements (covariance between distinct spins) can be reduced. In the naive and time-optimized protocols, where the system primarily occupies the all-up and all-down configurations, all spin covariances (even for non-neighboring spin pairs) are large where those configurations are equally populated. The fully optimized protocols disrupt this bistability and lower the total energy of many other configurations; this reduces spin covariance, with low covariance between neighboring spins and near-zero covariance between non-neighboring spins. The fully optimized 4D protocol, with additional controls, reduces covariance more than the 2D protocol.

The covariance between spins reflects the internal energy that couples their orientations; high covariance stems from ensembles dominated by low-internal-energy configurations with aligned spins, while low covariance results from the system accessing configurations with anti-aligned spins and hence higher internal energy. We, therefore, next investigate the energetic and entropic properties of instantaneous system distributions along the control protocol.

\begin{figure}[t] 
	\centering 
	\includegraphics[width=\linewidth]{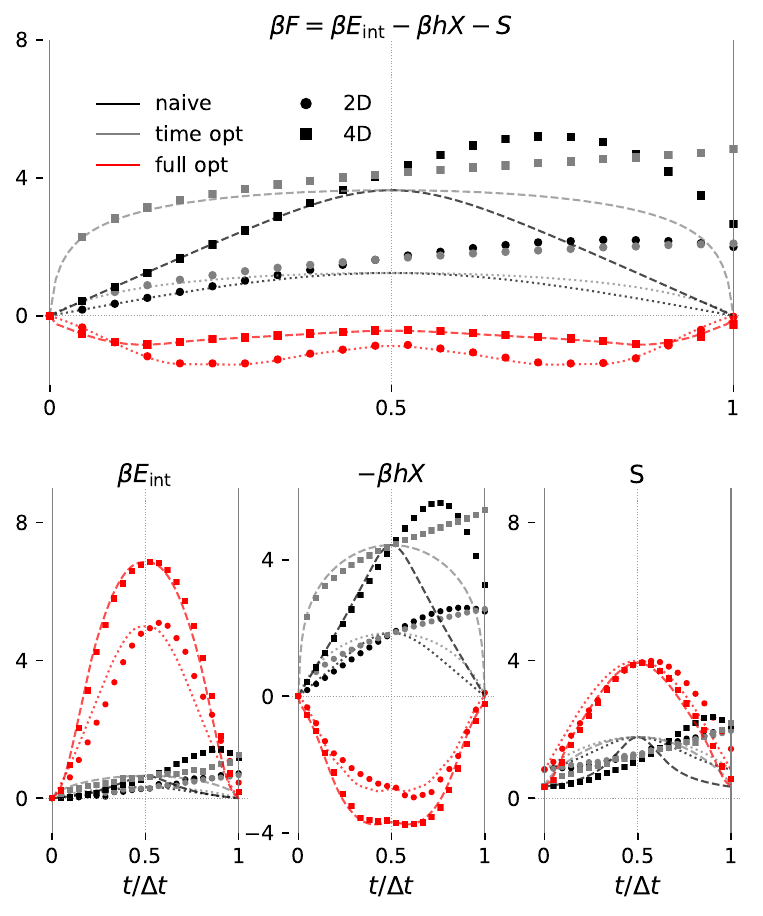}
	\caption{Free energy (top), and its components the mean internal energy~\eqref{eq:system_energy} (bottom left), mean external energy (bottom middle), and entropy (bottom right), of equilibrium (curves) and nonequilibrium (points) distributions, as a function of elapsed time $t$ along a moderate-duration ($\Delta t \approx \tau_{\rm rel}$) protocol of each type.}
	\label{fig:protocol_eqm}
\end{figure}

Figure~\ref{fig:protocol_eqm} shows the free energy along each protocol, as well as its component energies~\eqref{eq:control_energy} and entropy $S\equiv-\sum_{\{ \bm{\sigma} \}} p(\{ \bm{\sigma} \}) \ln p(\{ \bm{\sigma} \})$ for (equilibrium or nonequilibrium) probability distribution $p(\{ \bm{\sigma} \})$. The fully optimized protocols drive the system through distributions with high mean internal energy~\eqref{eq:system_energy}. This indicates significant population of high-internal-energy configurations (i.e., with anti-aligned spins), relative to naive and time-optimized protocols where the internal energy is relatively constant (consistent with the system primarily occupying the all-down or all-up configuration). These configurations with high internal energy are stabilized by stronger external fields, flattening the total-energy landscape relative to naive and time-optimized protocols. The fully optimized protocols also significantly increase the entropy, indicating many accessible system configurations, in contrast to naive and time-optimized protocols which primarily fluctuate near the all-up and all-down configurations. 

The flattened total-energy landscape and increased entropy during the fully optimized protocol results in relatively constant free energy, requiring comparatively low-magnitude quasistatic work throughout the protocol. In contrast, the naive and time-optimized protocols drive the system through a substantial free-energy barrier, requiring quasistatic work input to the system during the first half of the protocol and quasistatic work extraction during the second half.

\subsection{Heterogeneously inputting and extracting work, coinciding with system relaxation}

We now investigate how fully optimized protocols keep the free energy relatively constant (compared to naive protocols), by splitting total work into quasistatic and excess works by each control-parameter component as follows:
\begin{subequations} 
\begin{align}
    \mathcal{W} &= \sum_i \left[ \mathcal{W}^{\rm{qs}}_i + \mathcal{W}^{\rm{ex}}_i \right] \label{eq:work_breakdowna} \\
    &= \sum_i \left[ \int_0^{\Delta t} \md t \> \mathcal{P}^{\rm qs}_i(t) + \int_0^{\Delta t} \md t \, \mathcal{P}^{\rm ex}_i(t) \right] \label{eq:work_breakdownb} \\
    &\approx \sum_i \Bigg[ - \int_0^{\Delta t} \md t \, \dot{h}_i(t) \langle X_i \rangle^{\rm eq}_{\bm{h}(t)}     \label{eq:work_breakdownc} \\
    &\quad + \int_0^{\Delta t} \md t \sum_j \dot{h}_i(t) \zeta_{ij}(\bm{h}(t)) \dot{h}_j(t) \Bigg] \nonumber \ .
\end{align}
\end{subequations}
The final line uses the linear-response approximation~\eqref{eq:excess_work}. The quasistatic works of all components sum to the path-independent equilibrium free-energy difference between control-protocol endpoints, $\sum_i \mathcal{W}_i^{\rm qs}=\Delta F$; however, the quasistatic work of each component is path-dependent. 

\begin{figure}[t] 
    \centering
    \includegraphics[width=\linewidth]{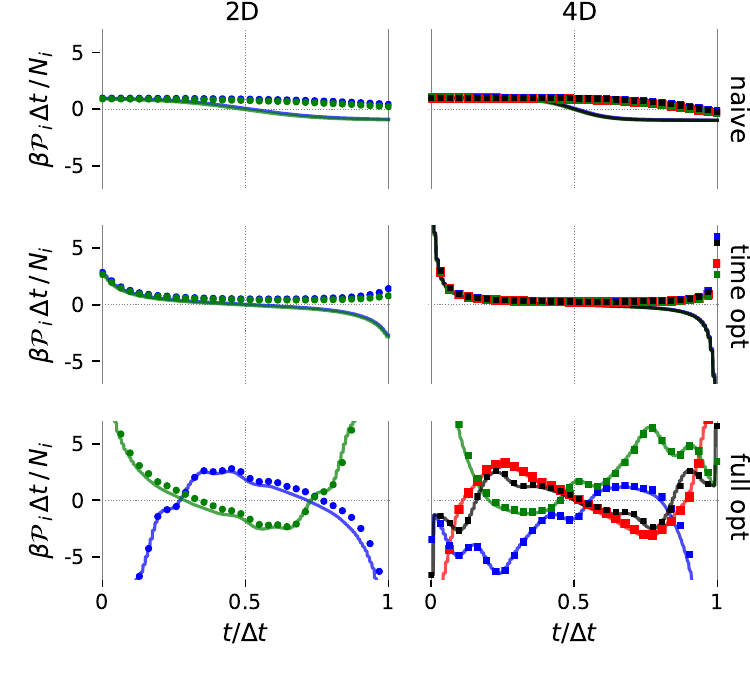} 
    \caption{Quasistatic power (curves) and finite-duration power [Eq.~\eqref{eq:work_breakdownb}, points] of each field (colors, see Fig.~\ref{fig:Ising_Model}) as a function of elapsed time $t$ along a moderate-duration ($\Delta t \approx \tau_{\rm rel}$) protocol of each type (see Fig.~\ref{fig:protocol_paths}). Power $\mathcal{P}_i$ is scaled by the number $N_i$ of spins controlled by field $i$.}
    \label{fig:work_power_protocol}
\end{figure}

Figure~\ref{fig:work_power_protocol} shows components of the quasistatic power $\mathcal{P}^{\rm qs}_i(t)$ and finite-duration power $\mathcal{P}_i(t)$, for moderate protocol duration ($\Delta t \approx \tau_{\rm rel}$). During naive and time-optimized protocols, each field inputs and extracts work equally, inputting work in roughly the first half of the protocol to increase the system's free energy and extracting it in the second half to reduce the free energy (Fig.~\ref{fig:protocol_eqm}). For finite duration, the mean conjugate force lags its equilibrium mean (Fig.~\ref{fig:protocol_mean_trajectories}), and the quasistatic work is not fully extracted in the second half of the protocol, resulting in excess work. The fully optimized protocols input and extract work differently through the different fields, with zero power for each component where either the field velocity is zero or the spin magnetization for that field is zero (i.e., controlled spins on average flip from down to up).

We examine the 2D fully optimized protocol, roughly dividing it into three stages according to sign changes of the power. Until $t/\Delta t \approx 0.3$, the green field is increased while the blue field is decreased (Fig.~\ref{fig:protocol_paths}), inputting power to the green spins to increase their mean magnetization (Fig.~\ref{fig:protocol_mean_trajectories}) while power is extracted from the blue spins (which do not significantly change mean magnetization), so that overall nearly no power is input to the system and the free energy is relatively constant (Fig.~\ref{fig:protocol_eqm}). Second (until $t/\Delta t \approx 0.7$), both fields are increased and work is input to the blue spins to increase their magnetization, while work is extracted from the green spins as they are stabilized in spin-up. Finally, the green field is reduced and work is input to the green spins, while increasing the blue field stabilizes the spin-up configuration of the blue spins and extracts work: the two fields act in concert to keep the free energy more constant. This elaborate and heterogeneous schedule of work input and extraction could reflect the system's energetic requirements to fluctuate into configurations with high internal energy: during naive and time-optimized protocols the system must wait for appropriate fluctuations from the environment to overcome the energy barrier; in contrast, the fully optimized protocols externally provide the appropriate energy to each spin as work, significantly decreasing the time required for system relaxation. The 4D fully optimized protocol shows similar features to 2D in the green and blue power, but events happen relatively earlier in green and later in blue. 

\begin{figure}[t] 
    \centering
    \includegraphics[width=\linewidth]{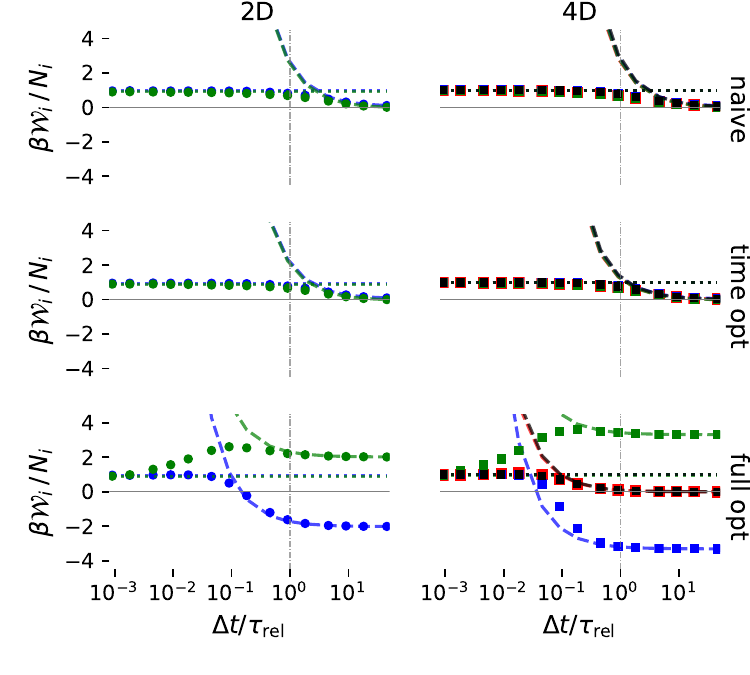} 
    \caption{Work~\eqref{eq:work_breakdowna}
    as a function of protocol duration $\Delta t$ for each field (colors, see Fig.~\ref{fig:Ising_Model}), for each protocol type (see Fig.~\ref{fig:protocol_paths}). Dotted lines: instantaneous work. Dashed curves: linear-response approximation. Protocol durations are scaled by reference spin-inversion relaxation time $\tau_{\rm rel}$ (Appendix~\ref{reference_relaxation_time}).}
    \label{fig:component_work_duration}
\end{figure}

Fig.~\ref{fig:component_work_duration} shows the work $\mathcal{W}_i$ by each field $i$ as a function of protocol duration, asymptoting to the instantaneous work ($-[h_i(t=\Delta t)-h_i(t=0)] \langle X_i \rangle^{\rm eq}_{\bm{h}(t=0)} \approx 1 \, \kT$ per spin) for short durations and reaching the (path-dependent) quasistatic work in the long-duration limit. For the naive and time-optimized protocols, each field does the instantaneous work until $\Delta t \approx \tau_{\rm rel}$ before crossing over to near-zero quasistatic values for longer durations. 

The fully optimized protocols input work via the green field and extract work via the blue field in the quasistatic limit. Work input from the green field increases from instantaneous to quasistatic values at short durations while work inputs from other fields remain near instantaneous values. When the duration is sufficiently long for the system to relax ($\Delta t \approx 0.1\tau_{\rm rel}$), the work from the remaining fields crosses over from instantaneous to quasistatic values, with the blue field extracting work from the system moderate-to-long duration. This suggests that work input by the green field early in the protocol is transduced through the system to be extracted by the blue field later in the protocol. Driving the green spins to flip early in the protocol initiates the transition mechanism, with further work input to red and black spins continuing the complete spin inversion by allowing the system to access high-internal-energy configurations, and final work extraction from blue spins stabilizing the system in the low-internal-energy all-up configuration. Providing work to specifically support this transition mechanism---rather than simply homogeneously inputting and extracting work to each spin type---keeps the system close to equilibrium for durations $\sim$10$\times$ shorter than naive and time-optimized protocols. 

\begin{figure}[t!]
    \centering 
    \includegraphics[width=\linewidth]{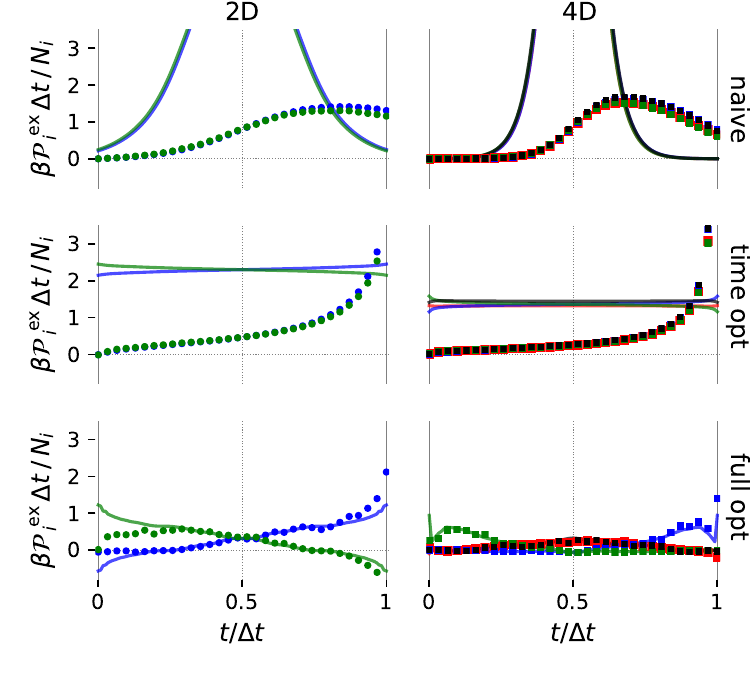}
    \caption{Exact excess power [points, Eq.~\eqref{eq:work_breakdownb}] and linear-response approximation [curves, Eq.~\eqref{eq:work_breakdownc}] for each field (colors, see Fig.~\ref{fig:Ising_Model}) as a function of elapsed time $t$ along a moderate-duration ($\Delta t \approx \tau_{\rm rel}$) protocol of each type (Fig.~\ref{fig:protocol_paths}).}
    \label{fig:excess_work_power_protocol}
\end{figure}

Figure~\ref{fig:excess_work_power_protocol} shows the excess power by each field for moderate duration ($\Delta t \approx \tau_{\rm rel}$), as well as linear-response approximations~\eqref{eq:work_breakdownb}. The excess power during naive and time-optimized protocols is essentially equal for each field, consistent with the field velocities (Fig.~\ref{fig:protocol_paths}) and lag of conjugate forces (Fig.~\ref{fig:protocol_mean_trajectories}) being equal for each component in these protocols. The linear-response approximation for the excess power during naive protocols peaks in the middle where friction is largest (Fig.~\ref{fig:2D_MWCP_friction_surface}). In time-optimized protocols, the linear-response approximation to the excess power is essentially the same for each field and constant, flattening the excess-power peak in naive protocols by reducing the protocol velocity in this region. The excess power at this moderate duration (significantly distant from equilibrium) unsurprisingly differs from the linear-response approximation for both naive and time-optimized protocols. Nevertheless, the excess power during the fully optimized protocol is significantly reduced relative to both naive and time-optimized protocols and agrees well with the linear-response approximation, indicating that the system remains much closer to equilibrium. 

\section{\label{sec:Discussion} Discussion}
The naive and time-optimized protocols (which follow the same path through multidimensional control-parameter space) change all fields simultaneously, biasing the total magnetization to drive the system from all-down to all-up. The barrier separating the energetically metastable configurations remains large throughout the protocol, and for moderate-duration protocols, the slow transitions between these endpoints prevent the system from relaxing to equilibrium resulting in significant lag and large excess work. The time-optimized protocol reduces velocity where the friction is large to allow more time for system relaxation, which for moderate-to-long durations reduces the excess work relative to the naive protocol.

In contrast, fully optimized protocols in multidimensional control-parameter spaces have more flexibility to manipulate the system's total energy~\eqref{eq:control_energy}; they use this flexibility to flatten the total-energy landscape through a heterogeneous schedule of work input and extraction through each spin (Figs.~\ref{fig:protocol_eqm} and~\ref{fig:work_power_protocol}), also speeding up system transitions from all-down to all-up. Both these equilibrium thermodynamic and kinetic properties are captured by the generalized friction tensor through the respective matrices of force covariance and integral relaxation time~\eqref{eq:friction_decomposition}. Both components of the friction play a role in improving our understanding of minimum-work control in multidimensional control-parameter spaces.

The force covariance matrix is a property of the equilibrium distribution without any reference to the system's dynamics. It quantifies the curvature of equilibrium free-energy in control-parameter space~\cite{Crooks2007}. For systems where the integral relaxation time is constant, the force covariance matrix replaces the generalized friction as the metric on control-parameter space that quantifies excess work, indicating that constant free-energy paths minimize work. Although the integral relaxation time is not constant throughout control-parameter space for our system [Fig.~\ref{fig:2D_MWCP_friction_surface}(c)], we still find that fully optimized protocols have relatively constant free energy relative to naive and time-optimized protocols (Fig.~\ref{fig:protocol_eqm}), a factor that helps to reduce work. 

We also see that off-diagonal elements of the force covariance matrix are significantly reduced relative to naive and time-optimized protocols (Fig.~\ref{fig:spin_covariance}), reflecting that the system accesses configurations which are otherwise inaccessible due to high internal energy. This has interesting connections to strategies for optimal inference of system parameters~\cite{Jiang2019}. There, a spin system is perturbed by external fields to new equilibrium ensembles that allow more efficient inference of system coupling constants (i.e., fewer samples are required to estimate parameters to a given precision), and the perturbation is updated using the Fisher information matrix for desired system parameters. Inference is optimal at field strengths that decorrelate spins, allowing the system to occupy otherwise-inaccessible configurations. Since the excess work during the protocol is known to affect the efficiency of free-energy estimation~\cite{Shenfeld2009,Kim2012,Blaber2020} and the force covariance matrix is proportional to the Fisher information matrix~\cite{Crooks2007}, it is not surprising that parallels in the design strategy exist. However, a key distinction is the dynamical nature of control protocols, where the relaxation kinetics between equilibrium distributions of successive control parameters is relevant; thus the importance of the generalized friction is reflected, which captures relaxation timescales via the integral relaxation time matrix.

The fully optimized protocols reveal a clear mechanism for spin inversion. First, work is input to green spins (initially energetically frustrated) causing them to flip, which together with direct work input by red and black fields drives neighboring red and black spins to flip. At the midpoint of the 4D fully optimized protocol, both green spins are up, red and black spins are evenly split between up and down, and blue spins are both down. Past the midpoint, work is extracted from red and black spins by increasing the corresponding fields to stabilize spins in the up orientation, which together with work input to neighboring blue spins causes blue spins to flip. Finally, work is extracted from blue spins to stabilize the system in the all-up configuration. 

Using this strategy, the system reaches the linear-response regime for durations $\sim$10$\times$ shorter than in the naive and time-optimized protocols (Fig.~\ref{fig:component_work_duration}). Excess work scales as $\Delta t^{-1}$ for long-duration protocols where the system is in the linear-response regime, so a heuristic strategy to minimize work is to find a path between configuration endpoints where the system relaxes quickly so that the system reaches the linear-response regime for shorter-duration protocols. The fully optimized protocols appear to facilitate such fast-relaxing paths in configuration space, reducing the time spent in the initial metastable basin before making a fast transition over the energy barrier. Earlier studies showed an empirical similarity between the minimum-work protocol and the minimum free-energy path~\cite{Rotskoff2017,Venturoli2009}, which represents the spontaneous (in the absence of driving) transition path between metastable mesostates. Our minimum-work protocols here also drive the system along a plausible spontaneous transition path: green spins flip first, followed by red and black spins, and finally blue spins. There may be a deeper connection between minimum-work protocols and spontaneous transition paths of the unperturbed system, which would allow extraction of kinetic information about transition paths from minimum-work protocols or use kinetic information to design efficient protocols. A more detailed quantitative comparison is necessary to fully elucidate this connection.

{\ML In this work, we assume throughout a particular set of control parameters that independently bias the energy of each spin set, since we expect advantages to accrue to distinct control of spins with different boundary conditions. In general, choosing a small set of relevant coordinates with which to drive the system is a non-trivial task~\cite{Peters2016}.
Recent advances in machine learning allow determination of small sets of coordinates that can drive the system across energy barriers, generally in enhanced-sampling 
contexts~\cite{Wang2021, Ma2005, Noe2013, Perez-Hernandez2013, Mardt2018, Tiwary2016a, Wang2019, Hernandez2018}.
Nonequilibrium driving protocols and enhanced-sampling methods can both probe the same space of control parameters that restrain the system and face similar convergence challenges related to system relaxation between states characteristic of distinct control-parameter values. We thus expect that the small coordinate sets identified by machine learning as fruitful for enhanced sampling would also serve as useful control parameters for minimum-work control.}

\section{\label{sec:Conclusion} Conclusion}
We have closely studied protocols designed to minimize work in driving magnetization inversion of a 2D Ising model, uncovering {\ML physically intuitive} principles of multidimensional control that reduce work relative to one-dimensional protocols. {\ML We characterize in detail the differences between naive, time-optimized, and fully optimized protocols and analyze the energetic and temporal aspects of the fully optimized protocols that significantly improve their performance relative to one-dimensional protocols.}
Designed multidimensional protocols take advantage of their more flexible control to avoid control-parameter regions of high friction (Fig.~\ref{fig:2D_MWCP_friction_surface}
{\ML, also seen in Refs.~\cite{Rotskoff2015} and~\cite{Gingrich2016}
}), decorrelate spins (Fig.~\ref{fig:spin_covariance}
), flatten the energy landscape thereby boosting the population in configurations with high internal energy (Fig.~\ref{fig:protocol_eqm}
), and heterogeneously input and extract work in concert with system relaxation timescales (Fig.~\ref{fig:work_power_protocol}).
This drives the system along a fast-relaxing path connecting the configuration endpoints, keeping the system closer to equilibrium and reducing resistance and hence work. It would be interesting to study multidimensional control in other model systems to critically assess the generality of these proposed design principles. 

\begin{acknowledgments}
This work was supported by the Natural Sciences and Engineering Research Council of Canada (NSERC) Canada Graduate Scholarships Masters and Doctoral (MDL), an NSERC Discovery Grant (DAS), and a Tier-II Canada Research Chair (DAS). Computational support was provided, in part, by WestGrid (www.westgrid.ca) and Compute Canada Calcul Canada (www.computecanada.ca). The authors thank Deepak Gupta and Steven Blaber (SFU Physics) for insightful comments on the manuscript.
\end{acknowledgments}

\section*{Data Availability Statement}
The data that support the findings of this study are available from the corresponding author upon reasonable request.

\appendix{}

\section{Numerical details} \label{computational_details}

\subsection{Calculation of friction tensor}

The friction matrix for control parameter $\bm{h}$ was estimated from time-correlation functions of pairs of conjugate forces~\eqref{eq:friction_tensor}. We used single-spin-flip Glauber dynamics~\cite{Glauber1963} to simulate four equilibrium trajectories, each of length $10^7 \md t$, where $\md t$ is a constant simulation time interval after which a random spin flip is attempted. Time-correlation functions were calculated between pairs of conjugate-force trajectories using fast Fourier transforms~\cite{Press2007} and integrated to lag time $2.5 \times 10^4 \, \md t$. The four independent estimates of friction at each control-parameter value were then averaged.

We calculated the 4$\times$4 friction matrix at grid points with discrete spacing $\md h_i = 0.2$ within the control-parameter space bounded by: $h_{\rm red} \in [-2.1,2.1]$, $h_{\rm blue} \in [-3.0,1.0]$, $h_{\rm green} \in [-1.0,3.0]$, $h_{\rm black} \in [-2.1,2.1]$. Since this control-parameter space has spatial symmetry, we doubled statistical power by averaging friction values for symmetrically equivalent control-parameter values: $(h_{\rm red},h_{\rm blue},h_{\rm green},h_{\rm black}) \to (-h_{\rm red},-h_{\rm green},-h_{\rm blue},-h_{\rm black})$.

\subsection{Design of control protocols}

{\ML 
Here we describe in more detail the methods used to design minimum-work protocols. 
		
The time-optimized protocols are effectively one-dimensional with a single field applied equally to all spins; these protocols have optimal velocity inversely proportional to the square root of the 1D friction coefficient $\zeta^{\rm 1D}(\bm{h}) \equiv \sum_{i,j} \zeta_{ij}(\bm{h})$. 
		
The fully optimized protocols are calculated using the string method~\cite{Rotskoff2017}.

Calculating both time-optimized and fully optimized protocols requires knowledge of the $4 \times 4$ friction matrix in relevant regions of control-parameter space.} \stkout{To calculate optimal protocols, we interpolated friction values and their gradients with respect to control parameters,} {\ML We interpolated each component of the friction matrix and its gradient (which is required for fully optimized protocols) with respect to each control parameter,} using {\ML four-dimensional} cubic splines~\cite{Walker2019,Walker2019b}, which solve piecewise cubic polynomials over the data array to ensure smooth first and second derivatives. {\ML Since components of the friction matrix vary several orders of magnitude throughout control-parameter space, we fit cubic splines to the natural logarithm of each matrix component and re-exponentiated interpolated values from this surface to recover the friction coefficient. We can take the natural logarithm since for our model each friction matrix component is positive. In low-dissipation regions at relatively large field values (top and left edges of the 2D control-parameter space in Fig.~\ref{fig:2D_MWCP_friction_surface}), noise causes the friction estimate at some values to become negative; we set these to a constant value of $10^{-5} \tau_{\rm rel}/ \kT$.}

To calculate the time-optimized protocols (Fig.~\ref{fig:protocol_paths}, middle), 
we calculated interpolated friction values at $121$ evenly spaced points along the fully naive protocol and then reparameterized these points to be evenly spaced in terms of the 1D friction $\zeta^{\rm 1D}(\bm{h}) \equiv \sum_{i,j} \zeta_{ij}(\bm{h})$. We \stkout{modified} {\ML used} the reparameterization scheme commonly used with the string method {\ML to keep string points evenly spaced~\cite{Maragliano2006}
with the modification that} \stkout{to measure} distances between discrete points {\ML are measured} using the friction metric~\cite{Rotskoff2015}.
{\ML This ensures constant excess power along this protocol (Fig.~\ref{fig:excess_work_power_protocol}).
}
	
To calculate the fully optimized protocols (Fig.~\ref{fig:protocol_paths}, bottom), 
we numerically solved the multidimensional Euler-Lagrange equation~\cite{Morin2008}
for excess work~\eqref{eq:excess_work}
using the string method, as developed in Ref.~\cite{Rotskoff2017}
{\ML This involves numerically solving the Euler-Lagrange equation for the protocol minimizing the excess work, by dividing the protocol into a discrete set of points along a ``string'' that runs between fixed protocol endpoints. We initialize this string as the naive protocol, where each point is evenly separated in Euclidean space. The string is updated from the $n$th to $(n+1)$th iteration by solving the set of linear equations for each component $h_{\alpha}$ at each discrete time $t$ along the string:
\begin{subequations}
\begin{align}
	&h^{n+1}_{\alpha}(t) - h^{n}_{\alpha}(t) = \Delta r \Bigg( D^2 h_{\alpha}^{n+1}(t) + \\ \nonumber
	&\sum_{ijk} [\zeta^{-1}]^n_{\alpha k}(t) D h^n_{i}(t) D h^n_{j}(t)  \left[ \partial_i \zeta_{kj}(t)-\frac{1}{2} \partial_k \zeta_{ij}(t) \right] \Bigg) 
\end{align}
\end{subequations}
where
\begin{subequations}
\begin{align}
    D h_{\alpha}(t) &\equiv \frac{h_{\alpha}(t+\delta t)-h_{\alpha}(t-\delta t)}{2 \delta t} \\
    D^2 h_{\alpha}(t) &\equiv \frac{h_{\alpha}(t+\delta t)+h_{\alpha}(t-\delta t)-2h_{\alpha}(t)}{\delta t^2} 
\end{align}
\end{subequations}
are finite-difference estimators of time derivatives for the string and $\delta t$ is the difference in the scaled protocol time between adjacent string points.
$\Delta r$ is a parameter controlling the size of the string update; we found that $\Delta r=10^{-4}$ for the 4D protocol and $\Delta r=10^{-5}$ for the 2D protocol allowed convergence of each string to a constant-excess-power protocol (required for the time-optimized and fully optimized protocols). Unlike Ref.~\cite{Rotskoff2017}
we did not reparameterize the string to keep points equally spaced; omitting this step allows us to simultaneously optimize the spatial and temporal aspects of the fully optimized protocols.}  
	
The algorithm requires evaluation of the friction matrix and its {\ML spatial} gradients with respect to control parameters at arbitrary control-parameter values, which are obtained from the spline fit. {\ML Ref.~\cite{Rotskoff2017}
assumes the integral relaxation time is constant and therefore the friction matrix can be replaced by the force covariance matrix; its derivative is simply the third cumulant of conjugate forces, $\langle \delta X_i \delta X_j \delta X_k \rangle$. This assumption greatly simplifies the calculation, but is not a good approximation for our system, as can be seen in Fig.~\ref{fig:2D_MWCP_friction_surface}
where the variation in the relaxation time throughout the 2D control-parameter space is on the same order of magnitude as the variation in force covariance, and therefore the variation of each element contributes significantly to the variation of the total friction.}
Optimization was carried out using $121$ discrete string points while the symmetry of the protocol was constrained, $(h_{\rm red}(t),h_{\rm blue}(t),h_{\rm green}(t),h_{\rm black}(t)) = (-h_{\rm red}(\Delta t-t),-h_{\rm green}(\Delta t-t),-h_{\rm blue}(\Delta t-t),-h_{\rm black}(\Delta t-t))$. 	
The 4D fully optimized protocol was passed through a Gaussian filter~\cite{Press2007}
with $\sigma=0.03 \Delta t$ to smooth noisy fields near protocol endpoints resulting from noisy friction estimates in low-dissipation regions. The smoothing affects the corresponding linear-response approximation to excess power in Fig.~\ref{fig:excess_work_power_protocol},
which visibly dips for green and blue fields close to the start and end of the protocol, respectively. However, the total excess power throughout the vast majority of the protocol is constant as expected.

\subsection{Protocol simulations}

To simulate the control protocol, we initialized a trajectory from the equilibrium distribution at $\bm{h}(t=0)$ and then calculated its spin-flip dynamics for duration $\Delta t$ while changing the control parameters according to the protocol. $N=5000$ repetitions of each driving protocol were used to compute nonequilibrium averages of power, work, spin magnetization, total energy, entropy, and free energy. Quasistatic properties are calculated from Boltzmann-weighted equilibrium distributions along the protocol.

\section{Calculation of reference relaxation time} \label{reference_relaxation_time}

Since we are interested in driving the system from all-down to all-up, we chose a reference relaxation time that reflects the timescale of this transition. The rate constant $k(\bm{h})$ for the reaction from all-down to all-up at equilibrium for control-parameter $\bm{h}$ is calculated from transition-path theory~\cite{Vanden-Eijnden2014}. This gives the mean number of all-down to all-up transitions observed per unit time, given that the system is in the metastable basin around the all-down configuration. We averaged this rate constant over control-parameter values along the 4D naive protocol to obtain an average rate constant $\Bar{k}$ for the spin inversion during this protocol: 
\begin{equation}
    \Bar{k} = \frac{1}{\Delta t} \int_{0}^{\Delta t} \md t \ k(\bm{h}(t)) \>.
\end{equation}
$\Bar{k}$ represents the mean number of transitions from all-down to all-up during a naive 4D protocol of duration $\Delta t$ (since the system begins near the all-down configuration, this conditioning is satisfied), assuming the system is in local equilibrium throughout the protocol. $\tau_{\rm rel}=\Bar{k}^{-1}$ then is the protocol time required to make on average one spin-inversion transition. For our system, $\tau_{\rm rel} = 1102 \, \md t$, where during each $\md t$ we attempt to flip one spin. The moderate-duration protocols shown in Figs.~\ref{fig:protocol_mean_trajectories}, \ref{fig:protocol_eqm}, \ref{fig:work_power_protocol}, and \ref{fig:excess_work_power_protocol} have duration $\Delta t = 1000 \, \md t \approx 0.91 \tau_{\rm rel}$.

%\bibliography{references}


\end{document}